\documentclass{PoS}
\usepackage{epsfig}
\usepackage{bm}
\usepackage{amssymb}
\usepackage{fontenc}
\usepackage{times}
\usepackage{mathptmx}
\usepackage{graphicx}

\title{The chiral phase transition for QCD with sextet quarks}

\ShortTitle{The chiral phase transition for QCD with sextet quarks}

\author{\speaker{D.~K.~Sinclair}%
\thanks{This research was supported in part by US Department of Energy
contract DE-AC02-06CH11357.}\\
HEP Division, Argonne National Laboratory, 9700 South Cass Avenue, Argonne,
IL, 60439, USA\\
        E-mail: \email{dks@hep.anl.gov}}

\author{J.~B.~Kogut
\thanks{Supported in part by a National Science Foundation grant 
NSF PHY03-04252.}\\
Department of Energy, Division of High Energy Physics, Washington, DC 20585,
USA\\
and\\
Department of Physics -- TQHN, University of Maryland, 82 Regents Drive, 
College Park, MD 20742, USA\\
        E-mail: \email{jbkogut@umd.edu}}

\abstract{QCD with 2 massless colour-sextet quarks is studied as a model of
Walking Technicolor. We simulate lattice QCD with 2 light color-sextet
staggered quarks at finite temperature, and use the dependence of the coupling
at the chiral transition on the temporal extent, $N_t$, of the lattice in 
lattice units to study the running of the bare lattice coupling with lattice 
spacing. Our goal is to determine whether this theory is QCD-like and `walks', 
or if it is conformal. If it is QCD-like, the coupling at the chiral transition 
should tend to zero as $N_t \rightarrow \infty$ in a manner controlled by
asymptotic freedom, i.e. by the perturbative $\beta$-function. On the other
hand, if this theory is conformal, this coupling will approach a non-zero 
limit in the $N_t \rightarrow \infty$ limit. We are extending our simulations
on an $N_t=8$ lattice to determine the position of the chiral transition with
greater accuracy, and are performing simulations on an $N_t=12$ lattice.}

\FullConference{XXIX International Symposium on Lattice Field Theory \\
		 July 10--16 2011\\
		 Squaw Valley, Lake Tahoe, California}

\begin{document}

\section{Introduction}

We are interested in extensions of the Standard Model with strongly-interacting
Higgs sectors. The most promising theories of this type are
Technicolor models \cite{Weinberg:1979bn,Susskind:1978ms}. 
Technicolor theories are QCD-like theories where the 
techni-pions play the r\^{o}le of the Higgs field, giving masses to the $W$ and
$Z$. Walking Technicolor theories, where the fermion content is such that the
coupling constant evolves very slowly -- `walks' -- and their extensions can
avoid phenomenological problems, which otherwise afflict Technicolor models
\cite{Holdom:1981rm,Yamawaki:1985zg,Akiba:1985rr,Appelquist:1986an}.

QCD with $1\frac{28}{125} \le N_f < 3\frac{3}{10}$ flavours of massless 
colour-sextet quarks is expected to be either a Walking or a Conformal field 
theory. (First term in the $\beta$ function is negative, second positive.) 
For $N_f=3$ conformal behaviour is expected. $N_f=2$ could, a  priori, exhibit 
either behaviour. We use unimproved lattice QCD with staggered quarks to 
simulate these theories. The RHMC simulation algorithm \cite{Clark:2006wp} is
used to allow us to simulate at values of $N_f$ which are not multiples of $4$.

We simulate $N_f=2$ thermodynamics in an attempt to distinguish whether it
is walking or conformal. Simulations on $N_t=4,6,8$ lattices show widely 
separated deconfinement and chiral symmetry restoration transitions, which
move to weaker couplings as $N_t$ is increased \cite{Kogut:2010cz,Kogut:2011ty}.
For the chiral transition, the separation between the $N_t=8$ and $N_t=6$
transitions is considerably less than that between the $N_t=6$ and $N_t=4$
transitions. 2-loop perturbation theory suggests that between $N_t=8$ and
$N_t=6$ the shift is consistent with asymptotic freedom, while between $N_t=6$
and $N_t=4$ the theory is strongly-coupled and the shift is due to quenched
evolution of the coupling. If this is correct, the deconfinement transitions
occur in the strong-coupling (quenched) regime for any $N_t$ we could
reasonably study. We are thus limiting our studies to the evolution of the
couplings at the chiral transition as $N_t$ is increased. Other groups are
studying QCD with 2 colour-sextet quarks using different lattice actions
\cite{Shamir:2008pb,DeGrand:2008kx,DeGrand:2009hu,DeGrand:2010na,%
degrand_lattice11,Fodor:2008hm,Fodor:2011tw,kuti_lattice2011}.
Most of these studies are of the zero-temperature behaviour of this theory. A
consensus as to whether this theory is QCD-like or conformal has yet to be
reached.

We are now extending our simulations to $N_t=12$ lattices and plan 
simulations on $N_t=18$ lattices, to check if this theory is indeed QCD-like,
or whether we are observing the approach to a bulk chiral transition,
expected for a conformal theory. (Our simulations are limited to the
neighbourhood of the chiral transition -- weak coupling.)
Since, however, 2-loop perturbation theory predicts that 
$\beta_\chi(N_f=12)-\beta_\chi(N_f=8) \approx 0.12$, we first need to
increase the number of $\beta$ values and statistics in the neighbourhood of 
the chiral transition, for $N_t=8$. We are also adding simulations at a
smaller quark mass, $m=0.0025$. These additional $N_t=8$ runs also provide
evidence that the $N_f=2$ chiral transition is second order.

\section{Simulations at $N_t=8$} 

We are extending our simulations on $16^3 \times 8$ lattices in the
neighbourhood of the chiral transition. Our quark masses are $m=0.0025$,
$m=0.005$, $m=0.01$, $m=0.02$, which will allow us to access the chiral limit.
Whereas our earlier simulations used $\beta$ values separated by $0.1$, we
have decreased this to $0.02$ close to the transition. Near this transition we
use 50,000-trajectory runs at each $(\beta,m)$ for $m=0.0025$,
20,000-trajectory runs for $m=0.005$ and $m=0.01$ and 10,000-trajectory runs
at $m=0.02$. We are currently increasing or plan to increase our run lengths
at each of these masses.

The chiral condensates for each mass decrease as $\beta$ increases. More
importantly, as $\beta$ increases, the mass dependence of these condensates
becomes more pronounced. The decrease in the chiral condensate with decreasing
mass is such that it does appear that it will vanish in the chiral limit for
$\beta$ sufficiently large. However, the $\beta$ dependence of 
$\langle\bar{\psi}\psi\rangle$ is sufficiently smooth at all the masses of
our simulations, that we would need a precise analytical form to perform
a believable chiral $m \rightarrow 0$ extrapolation to determine where it
vanishes. This we do not have. Hence we examine the (disconnected) chiral 
susceptibilities
\begin{equation}
\chi_{\bar{\psi}\psi} = \frac{V}{T}\left[\langle (\bar{\psi}\psi)^2 \rangle
                                   -(\langle \bar{\psi}\psi \rangle)^2\right]
\end{equation}
where $V$ is the spatial volume of the lattice and $T$ is the temperature.
$\bar{\psi}\psi$ is a lattice averaged quantity. Because we only have 
stochastic estimators for $\bar{\psi}\psi$ (5 per trajectory), we obtain 
unbiased estimators of $(\bar{\psi}\psi)^2$ as the products of 2 different
estimators of $\bar{\psi}\psi$ for the same gauge configuration. 

\begin{figure}[htb]
\epsfxsize=4in
\centerline{\epsffile{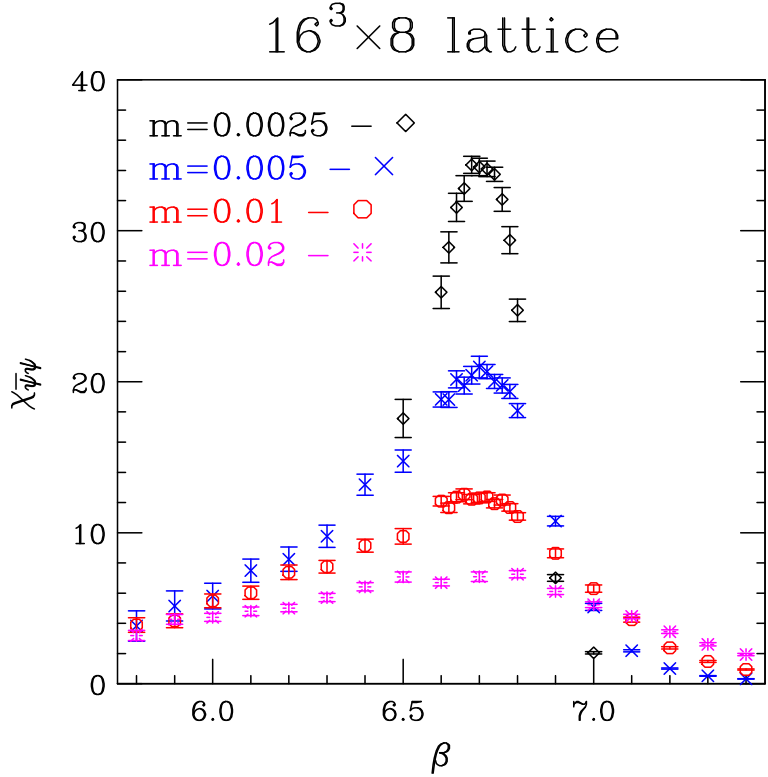}}
\caption{Chiral susceptibilities on a $16^3 \times 8$ lattice.}
\label{fig:chi8}
\end{figure}

The chiral susceptibility diverges at the chiral phase transition for zero
quark mass. At small but finite mass, it shows a clear peak which becomes
sharper as $m$ decreases. Extrapolating the position of said peak to $m=0$
yields $\beta_\chi$, the $\beta$ value of the chiral phase transition. 
Figure~\ref{fig:chi8} shows the chiral susceptibilities from our runs on
$16^3 \times 8$ lattices. What is clear from this plot is that the position of
the peak in the chiral susceptibility has very little dependence on the quark
mass $m$. It is this fact that allows us to extrapolate its position to $m=0$.
Our best estimate from the `data' presented here is $\beta_\chi=6.70(2)$.

\begin{figure}[htb]
\epsfxsize=4in
\centerline{\epsffile{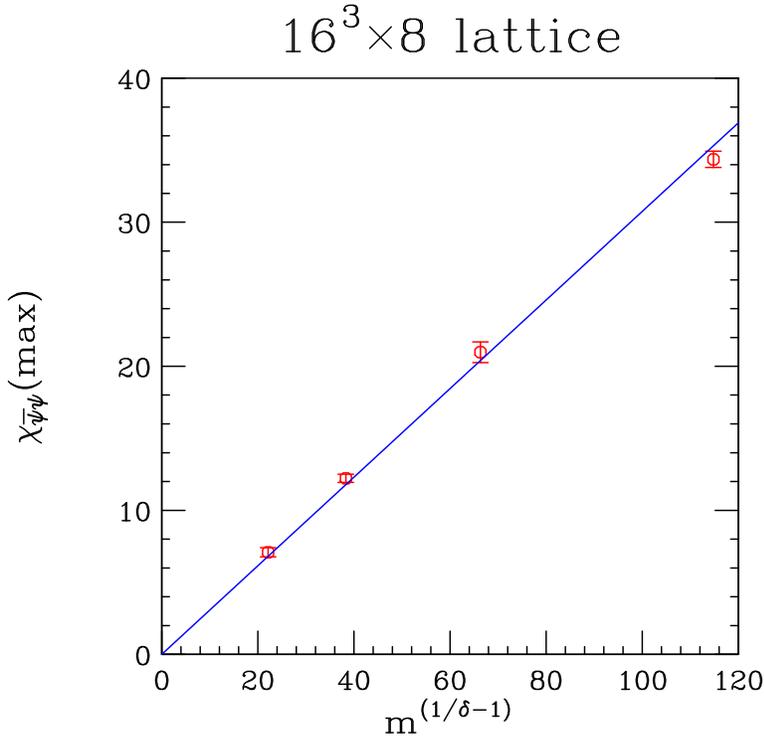}}
\caption{Peaks of the chiral susceptibilities for each mass as functions of
the scaling variable $m^{1/\delta - 1}$.}
\label{fig:scaling}
\end{figure}

Assuming that the chiral transition is a second order transition in the
equivalence class of the 3-dimensional $O(2)$ or $O(4)$ spin model, as 
expected for a finite temperature transition, we fit the peaks of the 
susceptibilities to the scaling form
\begin{equation}
\chi_{\bar{\psi}\psi}(max) = C m^{1/\delta - 1},
\end{equation}
with $\delta=4.8$. Figure~\ref{fig:scaling} shows that the fit is quite 
reasonable -- $\chi^2/DOF \approx 2.2$. This contrasts to the fit to the form 
for a first-order transition ($\delta=\infty$) which is very poor --
$\chi^2/DOF \approx 65$.

\section{Simulations at $N_t=12$}

Our $N_t=12$ simulations are being run on a $24^3 \times 12$ lattice for quark
masses $m=0.01$, $m=0.005$ and $m=0.0025$ and a range of $\beta$ values which
includes the chiral transition. We perform runs of 10,000 length-1 trajectories
away from this transition. Close to the transition ($6.6 \le \beta \le 6.9$)
the $\beta$ values are spaced by $0.02$, and we plan runs of 50,000 
trajectories for each $(\beta,m)$. The results we are reporting have
10,000--25,000 trajectories at each $(\beta,m)$.

\begin{figure}[htb]
\epsfxsize=4in
\centerline{\epsffile{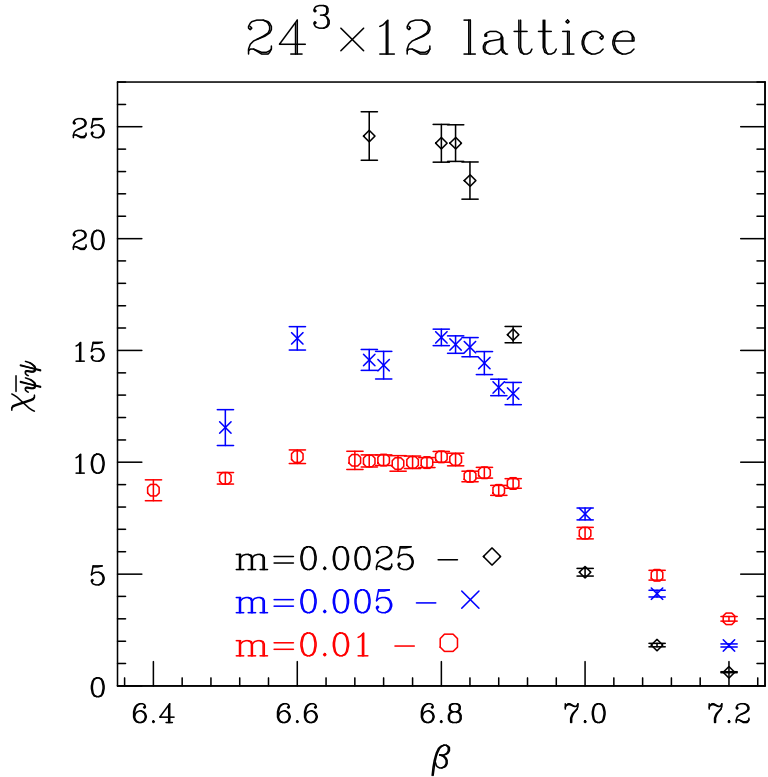}}
\caption{Chiral susceptibilities on a $24^3 \times 12$ lattice.}
\label{fig:chi12}
\end{figure}

Figure~\ref{fig:chi12} shows preliminary measurements of the chiral
susceptibilities obtained in these simulations. As one can see, the
susceptibility rises rapidly as $\beta$ is decreased from $\beta=7.2$ down to
$\beta$ just above $6.8$, for each of the 3 mass values. Below this each of
the graphs flattens out. This is obvious for $m=0.01,0.005$, and there is some
preliminary evidence for this at $m=0.0025$. From this preliminary `data' we
cannot estimate the position of the chiral transition. For this we will need
higher statistics and possibly a lower mass value.

\begin{figure}[htb]
\epsfxsize=4in
\centerline{\epsffile{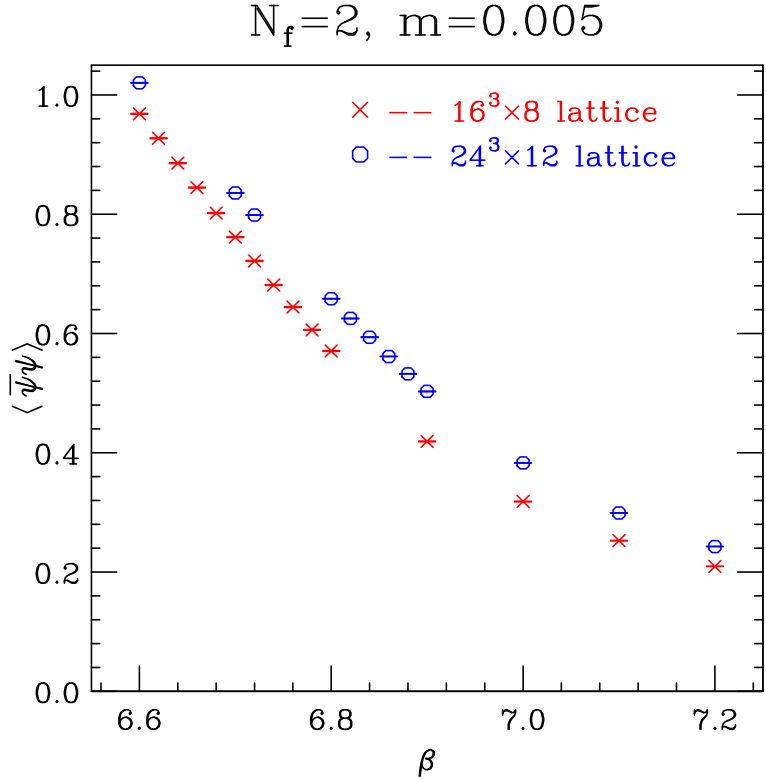}}
\caption{The $m=0.005$ chiral condensates on $16^3 \times 8$ and 
$24^3 \times 12$ lattices as functions of $\beta$.}
\label{fig:psi}
\end{figure}

Since we cannot as yet estimate the position $\beta_\chi$ of the chiral 
transition from the chiral susceptibilities, we perform a direct comparison
between the chiral condensates on the $16^3 \times 8$ and $24^3 \times 12$
lattices for the same mass. Ideally we should select the smallest mass
($m=0.0025$). However, we do not yet have enough $\beta$ values at $m=0.0025$
to make a good comparison. Hence we compare the chiral condensates at
$m=0.005$ on the two lattices. Figure~\ref{fig:psi} shows this comparison. The
$24^3 \times 12$ condensates are displaced to larger $\beta$ with respect to
the $16^3 \times 8$ condensates. However, the displacement appears to be less
than the $\approx 0.12$ predicted from the 2-loop Callan-Symanzik
beta-function.

\section{Discussion and conclusions}
We are simulating lattice QCD with 2 flavours of light staggered quarks at
finite temperatures on $16^3 \times 8$ and $24^3 \times 12$ lattices for 
$\beta$ close to the chiral transition. Our goal is to determine whether
the chiral transition for massless fermions approaches zero coupling
(infinite $\beta$) as $N_t \rightarrow \infty$ in a manner predicted by
the perturbative $\beta$-function, implying a QCD-like (walking) theory, or
if it approaches a finite non-zero value indicating a conformal field theory.
For $N_t=8$ we are able to determine the position of the chiral transition
$\beta_\chi$ with some accuracy ($\beta_\chi=6.70(2)$) from the peaks in the
chiral susceptibilities. The scaling of these peaks with mass is consistent
with a second-order phase transition in the $O(2)/O(4)$ universality class,
expected of a finite-temperature transition, and inconsistent with the 
first-order transition expected for a bulk transition. For $N_t=12$ our `data'
is still too preliminary to determine $\beta_\chi$ from the chiral
susceptibilities. However, direct comparison of the chiral condensates
suggests that $\beta_\chi(N_t=12) > \beta_\chi(N_t=8)$ as predicted by
asymptotic freedom for a finite temperature transition.

For $N_t=8$, we have also performed simulations on a $24^3 \times 8$ lattice
at a few strategically chosen values of $(\beta,m)$. Comparison of the
chiral condensates and chiral susceptibilities between the 2 spatial volumes
indicates that the finite volume effects are small. 

For $N_t=12$, we need more statistics at more values of $\beta$ for each mass,
to determine the peaks in the chiral susceptibilities. If the susceptibilities
remain too flat to be able to determine their peaks, we may need to simulate
at an even smaller mass ($m=0.00125$) to determine $\beta_\chi$. We will also
need to check that the finite volume effects are under control. We see no
reason why results as precise as those we obtained on $N_t=8$ lattices cannot
be obtained.

It is probable that simulations at even larger $N_t$s will be needed to
resolve the true nature of this theory. We are planning simulations at 
$N_t=18$ ($36^3 \times 18$ lattices). In addition we are planning some
zero temperature simulations in the chirally-restored phase.

\section*{Acknowledgements}
These simulations were performed on the Cray XT6 -- Hopper, the Cray XT4 --
Franklin and the Linux Cluster -- Carver/Magellan at NERSC, on the Cray XT5 -- 
Kraken at NICS under a TRAC allocation and on the LCRC's Linux Cluster -- 
Fusion at Argonne National Laboratory.


\begin{thebibliography}{99}


\bibitem{Weinberg:1979bn}
  S.~Weinberg,
  Phys.\ Rev.\  D {\bf 19}, 1277 (1979).

\bibitem{Susskind:1978ms}
  L.~Susskind,
  Phys.\ Rev.\  D {\bf 20}, 2619 (1979).


\bibitem{Holdom:1981rm}
  B.~Holdom,
  Phys.\ Rev.\  D {\bf 24}, 1441 (1981).

\bibitem{Yamawaki:1985zg}
  K.~Yamawaki, M.~Bando and K.~i.~Matumoto,
  Phys.\ Rev.\ Lett.\  {\bf 56}, 1335 (1986).

\bibitem{Akiba:1985rr}
  T.~Akiba and T.~Yanagida,
  Phys.\ Lett.\  B {\bf 169}, 432 (1986).

\bibitem{Appelquist:1986an}
  T.~W.~Appelquist, D.~Karabali and L.~C.~R.~Wijewardhana,
  Phys.\ Rev.\ Lett.\  {\bf 57}, 957 (1986).


\bibitem{Clark:2006wp}
  M.~A.~Clark and A.~D.~Kennedy,
  Phys.\ Rev.\  D {\bf 75}, 011502 (2007)
  [arXiv:hep-lat/0610047].


\bibitem{Kogut:2010cz}
  J.~B.~Kogut, D.~K.~Sinclair,
  Phys.\ Rev.\  {\bf D81}, 114507 (2010).
  [arXiv:1002.2988 [hep-lat]].

\bibitem{Kogut:2011ty}
  J.~B.~Kogut, D.~K.~Sinclair,
  [arXiv:1105.3749 [hep-lat]].


\bibitem{Shamir:2008pb}
  Y.~Shamir, B.~Svetitsky and T.~DeGrand,
  Phys.\ Rev.\  D {\bf 78}, 031502 (2008)
  [arXiv:0803.1707 [hep-lat]].

\bibitem{DeGrand:2008kx}
  T.~DeGrand, Y.~Shamir and B.~Svetitsky,
  Phys.\ Rev.\  D {\bf 79}, 034501 (2009)
  [arXiv:0812.1427 [hep-lat]].

\bibitem{DeGrand:2009hu}
  T.~DeGrand,
  Phys.\ Rev.\  D {\bf 80}, 114507 (2009)
  [arXiv:0910.3072 [hep-lat]].

\bibitem{DeGrand:2010na}
  T.~DeGrand, Y.~Shamir and B.~Svetitsky,
  Phys.\ Rev.\  D {\bf 82}, 054503 (2010)
  [arXiv:1006.0707 [hep-lat]].

\bibitem{degrand_lattice11}
  T.~DeGrand, Talk presented at Lattice2011, Squaw Valley, California (2011).

\bibitem{Fodor:2008hm}
  Z.~Fodor, K.~Holland, J.~Kuti, D.~Nogradi and C.~Schroeder,
  PoS {\bf LATTICE2008}, 058 (2008)
  arXiv:0809.4888 [hep-lat].

\bibitem{Fodor:2011tw}
  Z.~Fodor, K.~Holland, J.~Kuti, D.~Nogradi and C.~Schroeder,
  arXiv:1103.5998 [hep-lat].

\bibitem{kuti_lattice2011}
  J.~Kuti, Talk presented at Lattice2011, Squaw Valley, California (2011).

\end{thebibliography}
\end{document}